# Structural fingerprinting in the transmission electron microscope: Overview and opportunities to implement enhanced strategies for nanocrystal identification

Peter Moeck[1*] and Philip Fraundorf[2]

[1] Department of Physics, Portland State University, P.O. Box 751, Portland, OR 97207-0751, & Oregon Nanoscience and Microtechnologies Institute, U.S.A.
* Correspondence author e-mail: pmoeck@pdx.edu

[2] Department of Physics and Astronomy & Center for Nano Science, University of Missouri at St. Louis, MO 63121, U.S.A.

**Abstract.** This paper illustrates the prospective need for structural fingerprinting methods for nanocrystals. A review of the existing fingerprinting methods for crystal structures by means of transmission electron microscopy (TEM) which work for a single setting of the specimen goniometer is given. Suggestions are made on how some of these methods could be enhanced when nanocrystals and novel instrumentation are involved, i.e. when either the kinematic or quasi-kinematic scattering approximations are sufficiently well satisfied. A novel strategy for lattice-fringe fingerprinting of nanocrystals from Fourier transforms of high-resolution phase contrast transmission electron microscopy (HRTEM) images is briefly outlined. Nanocrystal structure specific limitations to the application of this strategy are discussed. An appeal is made to share the structural data of nanocrystals freely over the internet and infrastructure that would allow this sharing is mentioned. A shorter version of this paper has been accepted for publication in a special issue on the "Structure of Nanocrystals" of the "Zeitschrift für Kristallographie, International journal for structural, physical, and chemical aspects of crystalline materials".

Introduction

   The synthesis and processing of nanocrystalline structures has in many research laboratories resulted in materials with novel or improved physical, chemical, and biocompatible properties. When a large-scale commercial "nanocrystal powder-based industry" becomes a reality on the basis of such developments, quality, reproducibility, and affordability will be of essential interest to its customers [1]. Crucial to assessing quality and reproducibility of nanocrystal structures, sizes, and morphologies are characterization methods which are optimized for the nanometer size range.
   The well known powder X-ray diffraction method provides the standard means for the identification of a crystal structure [2] out of a range of candidate structures that are contained in a comprehensive database [3]. Identification processes such as this are generally referred to as structural fingerprinting. In powder X-ray diffraction fingerprinting, the three-dimensional (3D) crystal structure information is collapsed into a one-dimensional intensity profile plotted over the scattering angle or the magnitude of the reciprocal lattice vector. This ensures that the relative large abundance of structural 3D information can be used for the fingerprinting.

The position and relative heights of Bragg peaks in such diffractograms constitutes the information employable for structural fingerprinting. Certain technologically important nanomaterials, e.g., vanadium-oxide nanotubes [4], do, however, not give characteristic powder X-ray diffraction fingerprints by which the crystal structure could be identified out of a range of candidate structures from a comprehensive database. Such nanomaterials will, therefore, not become part of such a database.

The powder X-ray method [2] works best for crystal sizes in the micrometer range, but becomes less useful to useless for crystal structure identifications in the nanometer size range. This has for example been demonstrated experimentally for maghemite nanocrystals that were irradiated in a laboratory based diffractometer with a standard X-ray tube with a Cu target [5]. Similarly definitive in demonstrating the limited usefulness to uselessness of powder X-ray diffraction for structural fingerprinting in the nanometer size range are a series of simulations for $Ta_2O_5$ [6].

Nanometer crystal sizes have, on the other hand, exactly the opposite effect on the feasibility of the proposed lattice-fringe fingerprinting (with structure factor phase and amplitude extraction) of nanocrystals strategy because lattice fringes become visible over a wider angular range [7] and the combined weak phase object/kinematic[1] diffraction or phase object/quasi-kinematic[2] approximations of transmission electron microscopy (TEM) are reasonably well adhered to during the recording of the experimental data.

The goals of this paper are *(i)* to briefly outline a strategy for fingerprinting nanocrystals from crystal structure information contained in the Fourier transforms of high-resolution phase-contrast transmission electron microscopy (HRTEM) images [7-12] and *(ii)* to present it in the context of the other structural fingerprinting techniques that have been developed for TEM. The emphasis is thereby on the latter goal, i.e. on providing a rather comprehensive overview. On the basis of this overview, we will indicate where some of these techniques could be enhanced when (*i*) nanocrystals and/or (*ii*) novel instrumentation for either quasi-kinematic[2] electron diffraction or (quasi-)kinematic[1,2] highest-resolution phase-contrast imaging are involved. With necessity, such an overview can not be too detailed, but the reader may benefit from the many references to the key papers in the field.

It must be noted upfront that besides a rather comprehensive kinematic theory [7] and associated developments with respect to open-access crystallographic database support [8], there are at the moment only a few experimental proof-of-principle demonstrations [9-12] for the proposed strategy for fingerprinting nanocrystals from crystal structure information that is contained in the Fourier transforms of HRTEM images. The other ideas on how some of the established structural fingerprinting techniques in TEM could be enhanced for cases (*i*) and/or (*ii*) above are experimentally not proven, but seem to be quite reasonable (despite possibly turning out to become somewhat demanding on the experimental side). We present these ideas here in the hope that other researches will eventually take them up in order to develop robust techniques that allow for the structural fingerprinting of nanocrystals. We do this as a service to the scientific community because, as mentioned above, structural fingerprinting of nanocrystals can not be easily achieved [5,6] by the traditional powder X-ray diffraction techniques [2]. A common thread throughout all of these ideas is that structural fingerprinting in TEM can be made more discriminatory by including information that is to be extracted from experiments with fast electrons on nanocrystals on the basis of (quasi-)kinematic[1,2] scattering theories.

The Crystallography Open Database (COD) [13-16] (with currently more than 50,000 entries) and its mainly inorganic subset [17] are briefly mentioned in this context because they provide open access to comprehensive crystal structure information in the Crystallographic Information File (CIF) format [18]. CIF facilitates the communication of observational data and prospectively the exchange of crystallographic algorithms between the community of X-ray crystallographers and any other community that needs to solve crystallographic problems [19].

Because we will mention the extraction of structure factor phase angles (structure factor phases, not to be mistaken for "crystal phases") in the section on crystallographic image processing below, we will not speak about the identification of crystal phases from their characteristic "crystal phase fingerprints". Crystal phases (i.e. regions in space with homogenous physical and chemical properties) are referred to in this paper simply as "crystal structures" in order to avoid confusion with the structure factor phases.

Chemical fingerprinting information as obtainable in analytical TEMs by methods such as energy dispersive X-ray spectroscopy or electron energy loss spectroscopy may be used advantageously for the identification of crystal structures. Because we are in this paper only interested in giving an overview of methods that identify entities with periodic atomic arrangements from structural fingerprints in a TEM, chemical fingerprinting will not be discussed in this paper any further.

**Overview of existing crystal structure fingerprinting methods in TEM and strategies for their enhancement when nanocrystals are assessed**

The orders of magnitude stronger interaction of fast electrons with matter allows for the fingerprinting of much smaller crystals than those which can be identified (out of a range of candidate structures in a database [3]) via powder X-ray diffraction [2] with standard laboratory diffractometer. What appears as a very fine grained powder in X-ray diffraction with grain sizes in the sub-micrometer range is large enough for fingerprinting by means of single-crystal electron diffraction.

Kinematic[1] or quasi-kinematic[2] scattering approximations are likely to be applicable to sufficiently thin and small inorganic nanocrystals [20]. This allows not only for the solving of unknown structures by means of transmission electron microscope techniques [21-36], but also for proposals to enhance the existing crystal structure fingerprinting methods in TEM when such inorganic nanocrystals are assessed for identification purposes. The former is also known as (quasi-)kinematic-theory-based electron crystallography or structural electron crystallography, about which (the 1999 A. Lindo Patterson Award laureate) Douglas L. Dorset concluded: *„The success of ab initio structure analyses of inorganic crystals, based on electron diffraction intensities and, especially, on electron microscope images, is a surprising result, given the extreme pessimism expressed by many theoreticians … Most surprising is the success of high-resolution microscopy for structure determination."* [29]. One may argue on the basis of these quotes that if electron crystallography is feasible for sufficiently thin inorganic crystals, the much less sophisticated structural fingerprinting of inorganic nanocrystals for identification purposes on the basis of (quasi-)kinematic-theories should be feasible as well.

Many single-crystal electron transmission fingerprinting techniques only provide two-dimensional (2D) information on the crystal structures for any given setting of the specimen goniometer. Exceptions here include high-order Laue-zone convergent beam electron diffraction and multi-angle illumination techniques. Structural information in 3D is also present in oblique texture and electron ring diffraction pattern for a given setting of the specimen goniometer. This is because 2D projections of many differently oriented crystals will contribute to these patterns.

Employing more than one specimen goniometer setting for fingerprinting in TEM to obtain structural information in 3D from single crystals has been demonstrated, e.g. [37-44], but for brevity, this paper does not discuss most of these methods further. This is because fingerprinting of nanocrystals that involves more than one specimen goniometer setting is at the experimental level somewhat demanding and cannot easily be automated for the current generations of (not fully eucentric) double-tilt and tilt-rotation specimen goniometers. Manual single-crystal 3D fingerprinting methods can, on the other hand, be of help in deciding between a few candidate structures that resulted from database searches on the basis of information that was obtained by structural fingerprinting in 2D.

The (2D/3D hybrid) single-crystal method of ref. [44] works also for a single goniometer setting and employs a least squares fit to all of the electron diffraction spot positions in order to increase the precision with which the lattice parameters of nanocrystals can be extracted. This increase depends on the number of spots in the diffraction pattern in excess of the minimal set to solve the respective system of inhomogeneous linear equations and a threefold increase in accuracy has been demonstrated [45]. The method of ref. [44] has the additional advantage that a pair of electron diffraction spot patterns which are taken at known angular goniometer setting differences can be employed to make structural fingerprinting more discriminatory by including crystal lattice information in 3D. Moreover, the method is incorporated into a commercial software package (PhIDO/ELD)[3] for electron crystallography that also supports digitization and calibration of electron diffraction patterns [44].

The selected-area aperture in a modern TEM allows for the selection of crystal regions down to a few tenths of a micrometer. The much stronger interaction of fast electrons with matter, on the other hand, results frequently in multiple scattering. This necessitates the application of the dynamical diffraction theory to all problems that involve understanding diffracted intensities as a function of crystal thickness. For selected-area single-crystal 2D electron diffraction spot-pattern fingerprinting [45-48], the intensities of diffracted beams are generally neglected because dynamical scattering occurs and identifications of crystal structures are undertaken on the basis of the geometrical information that is contained in parallel-illumination electron diffraction spot patterns.

Single-crystal electron precession[4] diffraction [49,50], on the other hand, delivers quasi-kinematic[2] diffraction spot intensities for crystals that are up to several tens of nanometer thick [51] and wide. Employing the direct methods of X-ray crystallography (with atomic scattering factors for fast electrons), these quasi-kinematic diffraction spot intensities allow for the solving of structures from electron diffraction data [32-36]. The extraction of quasi-kinematic structure factor amplitudes for one setting of the specimen goniometer has for example been demonstrated in ref. [52].

It should, therefore, be possible to employ the experimentally obtained structure factor amplitudes as a component of structural fingerprinting in the TEM. Because the diffraction spot intensities are integrated by the electron precession movement, the projected 2D symmetry in the diffractogram could be employed as another component of structural fingerprinting. For this component of structural fingerprinting to work, the crystal would not need to be in an exact zone-axis orientation [51,53].

Compared to a conventional (selected-area) electron diffraction pattern from the same crystal, there are frequently many more diffracted spots in a precession electron diffraction pattern. This allows for least-squares fits to lager systems of inhomogeneous linear equations. The lattice parameters can, therefore, be extracted with a higher precision and the initial reciprocal lattice geometry based identification step of the structural fingerprinting be improved.

Kinematically forbidden electron diffraction spots may be present with very low intensities in electron precession diffractograms as a result of weak dynamical scattering [51,53]. The weak intensities of kinematically forbidden electron diffraction spots are further reduced by increasing the precession angle [53]. This dependency could be used to identify kinematically forbidden reflections. The quasi-kinematic diffraction spot intensities of electron precession also allow for the determination of space groups on the basis of spots from higher order Laue zones [53]. This could also be used for structural fingerprinting.

With respect to electron crystallography, it is advantageous that the so called "structure-defining" reflections (which are typically higher indexed and possess net-plane spacings from about 0.1 to 0.4 nm) fulfill the quasi-kinematic[2] diffraction approximations sufficiently well even for thicknesses on the order of 20 nm of crystals that are otherwise known to scatter fast electrons strongly dynamically [35]. For all of these reasons, electron precession[4] is likely to become the electron crystallographer's diffraction mode of choice for crystals that are up to several tens of nanometer thick [51] and wide.

It is typically observed that combining geometrical 2D information from a conventional single-crystal selected-area spot diffraction pattern, i.e. the two shortest reciprocal lattice spacings and the angle between these vectors, with information on the presence or absence of chemical elements within a crystal leads to unambiguous identifications of an unknown crystal structure out of a range of candidate structures in a comprehensive database [45-48]. (Such chemical information may be obtained from the same crystal in an analytical TEM by means of either energy dispersive X-ray spectroscopy or electron energy loss spectroscopy.)

Two of the structural fingerprinting techniques from conventional single-crystal (selected-area) spot diffraction pattern [45,46] rely on the Powder Diffraction File (PDF-2) [3] for the identification of a crystal structure out of a range of candidate structures. (The number after the acronym signifies the second format release of this database of the International Center for Diffraction Data (ICDD), which is a successor to the Joint Committee on Powder Diffraction Standard that was known under the acronym JCPDS).

Having been developed specifically to support structural fingerprinting from powder X-ray diffraction, the PDF-2 database [3], does, however, frequently not list the largest lattice plane spacings (so called d-values), which are often most diagnostic in electron diffraction data and lattice-fringe images. Unit-cell information from the Crystal Data database of the National Institute of Standards and Technology (NIST, formerly known as the National Bureau of Standards, NBS, [54]) is therefore used by two of the methods mentioned above [44,47]. The other one of these methods [48] relies on a database that resulted from collaborations between the NIST, Sandia National Laboratories, and the ICDD and is now known as the NIST Standard Reference Data Base 15 [55].

Employing nano-beam techniques [21], parallel-illumination electron diffraction spot patterns can be obtained from crystals as small as a few tens of nanometers. Even smaller nanocrystals can be illuminated individually by means of convergent beam electron diffraction. The high energy densities that are associated with convergent electron beams may, however, trigger structural transitions in nanocrystals. The structural fingerprinting techniques that are associated with convergent[5] electron beams (and the typically present Kikuchi lines) allow for orders of magnitude higher accuracies and precisions in the extracted information on the geometry of the reciprocal lattice [56-61]. These advantages of convergent electron beams for structural fingerprinting are, however, lost in the case of nanocrystals, where the diffraction disks are typically devoid of their fine structure [20].

Nanometer sized crystals can be fingerprinted on the basis of their parallel-illumination powder electron (ring) diffractograms [62,63] with support from either the Crystal Data [54] or the PDF-2 [3] databases. As in the case of powder X-ray diffraction fingerprinting [2], structural information in 3D is present, but there is a tendency of peak overlaps in electron diffraction ring patterns, especially for crystals of low symmetry (and when mixtures of small crystals with different structures are encountered [63]).

Employing one of the freely available computer programs from the community of X-ray crystallographers with atomic scattering factors for fast electrons, it has been demonstrated that the structure of a titania powder with an average grain size of about 7 nm could be refined [64] by means of the Rietveld method [65]. This result suggests that these nanocrystals diffracted either kinematically[1] or quasi-kinematically[2]. Two whole powder diffraction pattern decomposition methods are in common use for the extraction of the amplitudes of structure factors from X-ray powder diffractograms [66-68]. (For a review see ref. [69]). Because the Rietveld method has been employed successfully with atomic scattering factors for fast electrons to refine an inorganic nanocrystal structure [64], it should be possible to use these decomposition methods as well for electron diffraction ring patterns from nanocrystals that are small enough to scatter kinematically[1] or quasi-kinematically[2]. Good estimates of the unit cell parameters need to be available for either of these methods to work, so the electron diffraction ring pattern must be indexed and its reciprocal vector magnitudes measured rather accurately. The Le Bail structure factor extraction method [67,69] has the advantage that no assumptions about a model structure need to be made. The resultant structure factor amplitudes (including those that indicate systematic kinematic absences due to space group symmetry elements with translation components) are in any case very characteristic of a structure and should, therefore, enhance structural fingerprinting strategies from electron diffraction ring patterns.

With switching between the imaging and diffraction modes in a TEM, one can make sure that one fingerprints one crystal at a time by means of the parallel-illumination single-crystal electron diffraction techniques mentioned above. This is obviously no longer practical for electron powder diffractogram fingerprinting, although complementary dark-field images may be obtained by positioning a small objective aperture over a particular electron diffraction ring one at a time.

Identifications of crystal structures on the basis of the visual appearance of either oblique texture electron diffraction patterns [70] or of the fine structure within zone-axis orientation convergent beam electron diffraction (CBED) disk patterns [71,72] in the TEM have also been proposed. "Computer aided visual fingerprinting" of CBED disk pattern has been demonstrated where comparisons of simulated patters with experimental patterns are made automatically [61]. Because dynamical diffraction and crystal thickness effects are included into the simulations, the latter method is general and does not require zone-axis orientations [61].

Quite analogously, there has been a proposal to use the visual appearance of HRTEM structure images[6] from homologous series or isomorphic structures for fingerprinting purposes [73]. Such structure images need to be recorded at the Scherzer (de)focus from weak phase objects under strictly optimized conditions (preferentially including an objective-lens aperture that blocks all diffracted beam with spatial frequencies beyond the first zero crossing of the microscope's phase contrast transfer function) that make the appearance of the image essentially independent of the operation parameters of the TEM.

The visual appearance of HRTEM images is otherwise extremely sensitive to the instrumental operation parameters of the microscope, e.g. defocus, astigmatism, objective-lens aperture size, beam tilt, spatial and temporal coherence of the electron wave, and defocus spread [74]. In addition, the visual appearance of HRTEM images depends strongly on the thickness of the crystal and on any possible crystal tilt. (The former is the cause of a phenomenon that is known as "contrast reversal"). These crystal parameters need to be kept constant and must be as small as possible for structure-image fingerprinting to work. The resolution of structure images varies in a first approximation with the microscope's acceleration voltage and the spherical aberration constant of the objective lens so that different microscopes may produce structure images that appear visually different. HRTEM images can, on the other hand, be recorded from nanocrystals that are much smaller than those that are required for parallel-illumination single-crystal electron spot diffraction pattern fingerprinting.

Again quite analogously to the above mentioned "computer aided visual fingerprinting" of CBED disk patterns [61], it is customary to simulate HRTEM images on the basis of the dynamical diffraction theory, e.g. [75,76]. This is nowadays increasingly done over the Internet [77,78]. Because dynamical scattering effects, crystal thickness, crystal tilt, and the imaging conditions can be taken into account in such simulations, that method is also general. Due to the high sensitivity of the visual appearance of the HRTEM image to small changes in the actual imaging conditions, the "HRTEM image fingerprinting by simulation technique" is, however, quite time consuming. In addition, it is know that simulated images show approximately three times more contrast than experimental images [79], a phenomenon that is commonly referred to as "Stobbs factor".

In order to obtain a better figure of merit in comparing experimental HRTEM images with their calculated counterparts, it is customary to record so called "through focus series" of HRTEM images at defoci above and mainly below the Scherzer (de)focus with the objective aperture removed. Such defocus changes modify the phase contrast transfer function of the microscope's objective lens in a predictable manner, (which can for example be readily simulated with a freeware computer program [80].)

Increased underfocus (i.e. more negative than Scherzer (de)focus) emphasizes detail from higher spatial frequencies (smaller d-spacings if they are present) and, thus, "apparently seems to increase" the resolution of the microscope. Because the diffracted beams will be interfering with different objective lens induced phase shifts, the HRTEM image is then, however, no longer a directly interpretable structure image[6]. Nevertheless, this behavior might be used for lattice-fringe fingerprinting (both in the weak phase object/kinematic diffraction limit and somewhat beyond) in order to confirm or reject a match to a candidate structure from a comprehensive database.

While oblique texture and CBED disk pattern deliver 3D information for the structural fingerprinting process, the information in HRTEM structure images is restricted to 2D. With respect to the above mentioned fingerprinting from the fine structure of CBED disk pattern, it needs to be added that those crystals have to be sufficiently thick to give rise of dynamical diffraction effects [20]. At such thicknesses, the visual appearance of the corresponding HRTEM image will no longer be characteristic of a crystal structure as a result of violations of the weak phase object/kinematical diffraction approximations. Except in support of fingerprinting from the fine structure of CBED disks [81], no "visual atlas database" is readily available for the above mentioned fingerprinting methods in TEM that rely on visual appearances and comparisons.

Unfortunately, none of the above mentioned databases (that have previously been employed for structural fingerprinting in the TEM) [3,54,55] contains sufficient crystallographic information to implement the suggested enhancements on the basis of the kinematical scattering approximation. Such information would for example be the fractional atomic coordinates with respect to the unit cell and space groups (as needed, e.g., to calculate theoretical structure factor amplitudes for structural fingerprinting from electron precession data). Other databases such as the Inorganic Crystal Structure Database (ICSD) of the Fachinformationszentrum (FIZ) Karlsruhe [82], the PDF-4 of the ICDD [3], the newly released Pearson's Crystal Data (PCD) database of Material Phases Data System (MPDS), (Vitznau, Switzerland), the Japan Science and Technology Agency (JST), (Tokyo, Japan), and ASM International (the Materials Information Society, Materials Park, Ohio, USA), which is distributed by Crystal Impact (Bonn, Germany) [83] or the two comprehensive open-access databases Linus Pauling File of JST and MPDS [84] and Crystallography Open Database (COD) [13-17], would, therefore, need to be employed for the proposed enhancements.

**Crystallographic Image Processing and its possible usage for HRTEM lattice-fringe fingerprinting**

Crystallographic Image Processing (CIP) has its origins in the structural molecular biology community [85] and allows for the a posteriori correction of less than optimal recording conditions of HRTEM images of periodic weak-phase objects [86]. The application of CIP to HRTEM images of sufficiently thin inorganic crystals was first demonstrated by (the 1982 Chemistry Nobel Laureate) Aaron Klug in 1978 [85].

CIP enables electron crystallography from HRTEM images, i.e. the solving of structures of unknowns [22-30], to become both feasible and practical in the weak phase object/kinematic[1] diffraction limit approximations (and also somewhat beyond[2]). In essence, CIP results in reasonable approximations of the phases and amplitudes of the structure factors for the diffracted beams whose interference with the primary beam led to the formation of the HRTEM image. This is obtained by a Fourier analysis of the HRTEM image after it has been corrected for effects of the phase contrast transfer function of the objective lens (including possible two fold astigmatism, beam tilt, and crystal tilt).

Within the kinematic scattering approximation, the structure factors are directly proportional to the Fourier coefficients of the projected electrostatic potential that represents the HRTEM image. This potential and with it the outline of the projected atomic structure at the limited resolution of the HRTEM, i.e. at the very least the approximate location of the heaviest atoms [22,23], can be obtained by a Fourier synthesis of the experimentally determined structure factors. Because the effects of the phase contrast transfer function on the image are removed from the HRTEM image by CIP, there is no need to limit the resolution in these images to the Scherzer (or point-to-point) resolution of the TEM by an objective lens aperture. This is particularly important for TEMs with field emission electron guns where the information limit resolution is typically much higher than the Scherzer resolution.

A crucial component of CIP is the determination of the most probable projected crystallographic symmetries in an experimental HRTEM image. This is done on the basis of the knowledge that only 17 plane space groups exist, that the coefficients of any Fourier transform of an image must posses one of these symmetries, that Friedel's law applies, and that each of these space groups possesses its own unique set of theoretical phase angle relations and restrictions. On the basis of the relative agreement or disagreement of these theoretical phase relations and restrictions to their experimental counterparts, a figure of merit is defined that aids the determination of the most probable projected symmetry of the crystal under investigation. The procedure may involve the testing of all 17 plane space groups for many incrementally neighboring positions within a projected unit cell in an HRTEM image. This projected symmetry determination also ensures that the origin of the Fourier transform is fixed to the projected origin of the corresponding most probable three-dimensional space groups.

As mentioned above, other important components of CIP are the corrections for the effects of the phase contrast transfer function on the amplitudes of the Fourier coefficients of the projected electrostatic potential and the imposing of the projected symmetry on these amplitudes. The latter is equivalent to removing possibly existing crystal tilt a posteriori. CIP can nowadays be performed very efficiently on a desktop computer with the comprehensive software package CRISP[3].

Douglas L. Dorset concluded with respect to CIP that …"*when the measured intensity distribution of an experimental image is Fourier-transformed to find estimated crystallographic phases (imposing further constraints due to choice of unit cell origin and the space-group symmetry operators), it is not yet clear what amount of dynamical scattering can take place before these derived phase terms are completely meaningless.*" [29]. Given that frequently a useful structural model results from CIP of HRTEM images, these structural models should surely enhance structural fingerprinting of nanocrystals on the basis of candidate structures from a comprehensive database. The experimental intensity data and (quasi-)structure images[6] need to be close to (quasi-)kinematic limits, since it *"is always possible, even easy, to collect intensity data that **cannot** be analyzed by conventional phasing methods or to record high-resolution images where the resemblance to any known structure is not at all obvious."* (Dorset, [28]).

If the thickness of a nanocrystal exceeds the limit set by the validity of the weak phase object/kinematical approximation[1], it may still be sufficiently thin so that the pseudo-weak phase object approximation [87,88] or other quasi-kinematic approximations[2] might be valid. HRTEM imaging within the pseudo-weak phase object approximation means in essence that an isomorphic structure of a hypothetical approximant is imaged linearly instead of the real crystal structure. Dynamical diffraction effects result for this isomorphic hypothetical approximant in essence in a replacement of the lighter atoms by heavier atoms and of the heavier atoms by lighter atoms. The atomic coordinates, however, remain the same so that the structure factor phases of the isomorphic hypothetical approximant should be similar to their counterparts in the real crystal structure.

The structure factors and most importantly their phases are characteristic of the atomic arrangement for any given structure and must, therefore, be quite suitable for structural fingerprinting purposes. Structure factor phases can typically be extracted by means of CIP from experimental HRTEM images with an accuracy of approximately 20°. In addition, the phases are essentially unaffected by crystal tilts.

The absolute values of the structure factor amplitudes, which one can extract by means of CIP from HRTEM images, are less accurate. Their usage for structural fingerprinting purposes is, therefore, limited. Crystal tilt affects the structure factor amplitudes strongly and the phase contrast transfer function attenuates these amplitudes in addition in a nonlinear manner. Although these effects can be corrected by means of CIP, dynamical diffraction contributions for thicker crystals may render the values of the structure factor amplitudes meaningless for structural fingerprinting purposes. When CIP has been employed and dynamical diffraction contributions are not too strong, it may be possibly to infer space group information from the weakest structure factor amplitudes that would in the kinematic limit be zero. That space group information may then in turn be used for structural fingerprinting from HRTEM images.

Because CIP employs classical crystallography concepts (that are strictly speaking only valid for infinitely many and strictly precise repeats of structural entities in 3D), the structural perfection of the nanocrystal under investigation has to be reasonably high for the whole procedure to work satisfactory. It is found in practice that low densities of one and two dimensional defects can be tolerated. Because only some $10^2$ to $10^3$ unit cells are needed to employ CIP, the procedure should work for many nanocrystals.

For HRTEM images that are recorded under axial multiple beam imaging conditions [75] (with the microscope's magnification properly calibrated and its possible radial magnification dependency across the selected field of view corrected [23]), lattice constants can be derived from nanocrystal lattice fringes with accuracies on the order of 0.1 % [89-92]. Obtainable statistical precisions are for lattice constants on the order of 0.2 % [90] and for the (intersecting) angles between reciprocal lattice vectors (net-plane normals) on the order of 0.2° [91]. For the very smallest of nanocrystals, e.g. those with a diameter of about 2 nm, the statistical precisions are up to one order of magnitude worse [89] and, thus, comparable to those that can be obtained from single-crystal electron diffraction spot patterns form much larger crystals.

On the basis of the above, a novel strategy of structural fingerprinting from Fourier transforms of HRTEM images could be based on a combination of the results of CIP with geometrical information on the reciprocal lattice. Both the geometric information on the reciprocal lattice and the structure factor phases can be extracted from HRTEM images rather accurately and precisely. This information is also rather characteristic of a crystal structure and shall, therefore, aid the identification of unknowns out of a range of candidate structures from a database. Following the classical search/match paradigm, one could start the search for possible matches in a database on the basis of the extracted reciprocal lattice information. In the matching step, the structure factor phases should be used in order to decide between the few candidate structures that resulted from the search procedure.

As already mentioned above, none of the comprehensive databases [3,54,55] that have traditionally been employed for structural fingerprinting in the TEM from electron diffraction pattern contains information that allows for the calculation of structure factors for high energy electrons. The Crystallography Open Database (COD) [13-16] and its mainly inorganic subset [17] are, on the other hand, suitable for this purpose because they provide comprehensive crystallographic information in open access in the Crystallographic Information File (CIF) format [18].

Crystallographic software will need to be written in support of lattice-fringe fingerprinting. If based on the COD entries, this software will be able to take advantage of the CIF concept where CIF data are supported by CIF dictionaries in computer readable form that define a range of crystallographic concepts algorithmically from the most basic crystallographic information in the database entries [19]. Such software has been written recently in support of lattice-fringe fingerprinting [8,9]. The exemplary output of that software, so called "lattice-fringe fingerprint plots" in both the kinematic and dynamical diffraction limit, is given in Figure 1.

The so called "interfringe angle", i.e. the angle under which lattice fringes intersect in HRTEM images, is plotted in such plots against the reciprocal lattice vector magnitude [7]. While there are two data points in lattice fringe finger-print plots for crossed fringes with different spacings, the crossing of two symmetrically related fringes results in just one data point (because the latter possess by symmetry the same spacing). More elaborate lattice fringe fingerprint plots may contain in the third dimension histograms of the probability of seeing crossed lattice fringes in an ensemble of nanocrystals. The equations for calculating such probabilities for an ensemble of randomly oriented nanocrystals are given in ref. [7]. Structure factor phases and amplitudes could be added in a forth dimensions. Instead of employing higher dimensional spaces, one could also stick with two-dimensional displays such as Fig. 1 and add to each data point a set of numbers that represent additional information that will make structural fingerprinting from HRTEM images more discriminatory. Note that atomic scattering factors for fast electrons are available in parameterized and computer readable form [76] so that structure factors could be calculated straightforwardly for HRTEM-image based structural fingerprinting purposes.

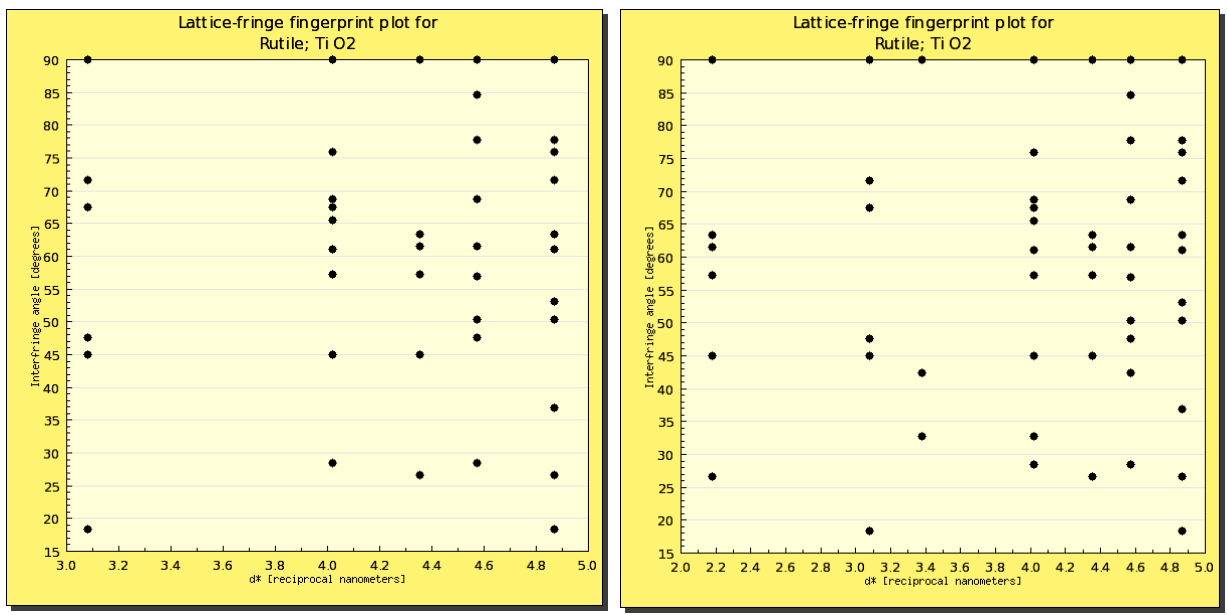

**Fig. 1:** Basic theoretical lattice-fringe fingerprint plots for rutile (TiO$_2$) for a microscope with a Scherzer resolution of 0.19 nm, form ref. [8]; (left) in the kinematic diffraction limit (right) in the dynamic diffraction limit. Such plots can be calculated over the Internet on the fly from the CIF format entries in the mainly in the mainly inorganic subset of the COD [17].

Because it would be based on HRTEM images, the proposed lattice-fringe fingerprinting strategy would have the advantage over diffraction pattern based fingerprinting that the nanocrystal size distribution (and morphology [43] when more than one goniometer setting is involved) could be assessed simultaneously. This ability would be of importance for prospective industrial customers of a "nanocrystal powder-based industry" [93].

The solving of materials science problems by means of TEM is currently undergoing the lens-aberration-correction revolution, e.g. [94]. Reliable spatial information down to the sub-Ångström length scale can nowadays be obtained in both the parallel illumination and the scanning probe (STEM) mode[7]. Objective-lens aberration-corrected TEMs and condenser-lens aberration-corrected STEMs [95] in the bright field mode allow for sufficiently thin crystals the retrieval of the Fourier coefficients of the projected electrostatic potential down to the sub-Ångström length scale and, thus, represent a novel type of crystallographic instrument in support of both image-based electron crystallography [22-31] and nanocrystallography [7-12,39,41,43]. Aberration corrected TEMs and STEMs are also expected to have an impact on the feasibility of electron crystallography on the basis of electron precession diffraction data [96].

The higher the point-to-point resolution in an aberration-corrected TEM is, the lower will in principle be the lateral overlap of the electrostatic potentials from adjacent atomic columns and the more zone axes will be revealed by crossed lattice fringes in structure images. Note that the relation between resolution and visibility of zone axes is typically strongly superlinear [7]. In addition, it is well known that significant excitation errors, as typical observed for nanocrystals that are oriented randomly on a TEM grid, (i.e. up to a few degrees away from lower indexed zone axis with respect to the electron beam [7],) effectively extend the thickness range over which the scattering of fast electrons may be approximated as kinematic[1] or quasi-kinematic[2]. Since aberration corrected TEM will eventually deliver sufficient resolution so that crossed lattice fringes can be discerned for almost all lower indexed zone axes for nearly all materials, the insensitivity of CIP [85,86] of HRTEM images to exact zone axis orientations should enhance the feasibility of the proposed structural fingerprinting strategy.

For a well-aligned aberration-corrected microscope, HRTEM images could be automatically recorded at a single setting of the specimen goniometer. Sub-Ångström spatial resolution would ensure that crossed lattice fringes would be visible for virtually all nanocrystals on a TEM support grid. These nanocrystals may possess either a texture or may be randomly oriented. CIP could be used to extract both the phases of the Fourier coefficients of the projected electrostatic potential (which are negligibly affected by crystal tilts) and the structure factor amplitudes. In short, a reliable and highly distinctive structural fingerprinting strategy could be run automatically on aberration-corrected microscopes.

When this structural fingerprinting procedure is automated, mixtures of nanocrystals could be analyzed both qualitatively and quantitatively. Since each nanocrystal would be identified individually and thousands of nanocrystals could be processed automatically, the detection limits for crystal structures could readily be pushed to levels that are superior to those of traditional powder X-ray diffraction fingerprinting.

**Nanocrystal specific limitations of the proposed lattice-fringe fingerprinting strategy**

The lattice constants of inorganic nanocrystals with diameters between ten and one nanometer may be contracted or expanded by a fraction of a tenth of a percent e.g. [97], a fraction of a percent, e.g. [98], up to a few percent at worst (for the very smallest crystals) due to free-energy minimization effects in the presence of a large surface areas [99] or due to reduced lattice cohesion [97]. For non-cubic nanocrystals, these lattice constant changes may result in changes of the angles between net planes (and, therefore, lattice fringes in HRTEM images) on the order of a tenth of a degree to a few degrees for some combinations of lattice fringes while other combinations may be negligibly affected. (As mentioned above, the average lattice parameters of nanocrystals can be obtained from a Rietveld refinement of electron diffraction powder data [64] and these averages could then become part of databases that are used to support the lattice fringe fingerprinting.)

To account for these possible size dependent lattice distortion effects of nanocrystals, lattice-fringe fingerprinting identifications of unit cell geometries might be based on the concepts of normalized reduced cells [100] and perturbation stable cells [101]. For the latter, possible distortions in the inter-axial angles of reduced unit cells are omitted in search/match procedures and all metrically similar lattices can be found [101].

It is also known that some nanocrystals possess surface and near surface areas that are highly distorted or relaxed with respect to the bulk crystal structure. (Byproducts of such distinct surface structures are X-ray powder diffraction patterns that are no longer characteristic of the crystalline bulk core [102]. Relaxed surface structures may have been observed experimentally by electron crystallography for colloidal $Pd_3P$ nanocrystals [26], because the CIP analysis of HRTEM images excluded the central parts of the nanocrystals[2]. Anatase ($TiO_2$) nanocrystals of less than about 2 nm may not possess a core region that corresponds to the bulk lattice structure [103]. Lattice-fringe fingerprints from such nanocrystals will, therefore, be quite challenging.

Another potential problem in HRTEM imaging of very small nanocrystals arises from the influence of support films or matrices [104] in which the nanocrystals might be embedded. These typically amorphous materials may preclude the lattice-fringe fingerprinting of very small nanocrystals by making sub-nanometer crystals effectively invisible.

Finally, there are "crystallographically challenged materials" [105,106] and "intercalated mesoporous materials" [107], i.e. tens of nanometer sized entities with a well defined atomic structure over small length scales that can be described by a relative large crystallographic unit cell with a low symmetry. The above mentioned vanadium oxide nanotubes [4] may serve as an example for the former class of materials. Both of these materials lack crystallographic long-range order. Significant structural distortions that might be considered as classical defects or nanocrystal specific defects to the average structure may be present to such an extent that it may make little sense to consider the disorder as a defect away from the ideal structure. In short, the deviations from the perfect atomic structure might be rather severe in these materials but remnants of the crystallinity might still be present. While mesoporous materials have been solved from HRTEM images by electron crystallography [108] and may, therefore, also be straightforwardly fingerprinted, the same needs to be demonstrated for crystallographically challenged materials. Such work is in progress and will be reported elsewhere for $V_2O_5$ nanotubes that have been both described in ref. [4] and also grown at Portland State University.

**Appeal to share nanocrystallographic structure data in open access**

Challenges to determine the atomic structure at the nanometer scale are best met by multidisciplinary teams working with complementary methods on either the same sample or the same batch of samples [107]. It was recently proposed that the data generated by these concerted efforts should be shared in a common computational global network [107] so that theorist can also make their unique contributions on the basis of the latest experimental data. The CIF format would facilitate interactions with theorist greatly as it is both well documented [18] and in widespread use amongst the experimentalist. Following these ideas and utilizing the COD model [13-16], an internet based open-access crystallographic database that collects entries on the structure of nanocrystals in this format [109] has recently been set up. The readers of this paper are encouraged to send their published nanocrystal structure results per email attachment to the first author of this paper so that they can be expressed in CIFs and made openly accessible. Ready made CIFs may in the future be uploaded directly.

Perceived copyright issues should not delay better communications within the emerging community of "nano-crystallographers" because "*facts*" [16] and "*ideas or data*" [110] can not be copyrighted under American [16] or International Law (as stipulated by the World Intellectual Property Organization [111] in the "Berne Convention for the Protection of Literary and Artistic Works" [112]). The "*expression incident to factual reporting*" [16] and "*original, tangible forms of expression*" [110] may, on the other hand, be subject to copyrights.

In CIF based open-access databases, such expressions are in the CIF format. The CIF format itself (and its underlying Self-defining Text Archive and Retrieval (STAR) structure) are owned by the International Union of Crystallography (IUCr), who "*will not permit any other organization to "capture" STAR or CIF and try to ransom it back to the community* [113]". The IUCr also asserts that "*if you are putting out your CIF or STAR compliant application to the world for free, we are not going to ask you to start charging money for it so that you can pay the IUCr a license fee* [113]".

**Summary**

Because an industrial scale need for structural fingerprinting of nanocrystals is emerging and since the current X-ray diffraction based structural fingerprinting methods do not satisfy this need, new transmission electron microscopical methods need to be developed. The existing fingerprinting methods for crystal structures by means of transmission electron microscopy which work for a single setting of the specimen goniometer were reviewed to aid such developments. Suggestions were made on how some of the existing structural fingerprinting methods could be enhanced when nanocrystals and/or novel instrumentation for either quasi-kinematic electron diffraction or highest-resolution phase-contrast imaging are involved. A novel strategy for lattice-fringe fingerprinting of nanocrystals from Fourier transforms of high-resolution phase contrast transmission electron microscopy images has been mentioned along with nanocrystal structure specific limitations to the application of this strategy.

**Acknowledgements**

This research was supported by an award from Research Corporation. Support was also provided by the Office of Naval Research, the Oregon Nanoscience and Microtechnologies Institute, the NorthWest Academic Computing Consortium, as well as by "Faculty Development", "Faculty Enhancement", and "Internationalization" Awards by Portland State University.

# References

[1] Saltiel, C.; Giesche, H.: Needs and opportunities for nanoparticle characterization. J. Nanopart. Res. 2 (2000) 325-326.

[2] Hanawalt, J. D.; Rinn, H. W.: Identification of crystalline materials: Classification and use of X-ray diffraction patterns. Ind. Eng. Chem. Anal. Ed. 8 (1936) 244-247.

[3] Faber, J.; Fawcett, T.: The Powder Diffraction File: present and future. Acta Cryst. B 58 (2002) 325-332.

[4] Petkov, V.; Zavalij, P. Y.; Lutta, S.; Whittingham, M. S.; Parvanov, V.; Shastri, S.: Structure beyond Bragg: Study of $V_2O_5$ nanotubes. Phys. Rev. B 69 (2004) 85410-1-85410-6.

[5] López-Pérez, J. A.; López Quintela, M. A.; Mira, J.; Rivas, J.; Charles, S. W.: Advances in the Preparation of Magnetic Nanoparticles by the Microemulsion Method. J. Phys. Chem. B 101 (1997) 8045-8047.

[6] Pinna, N.: X-Ray diffraction from nanocrystals, Progr. Colloid. Polym. Sci. 130 (2005) 29-32.

[7] Fraundorf, P.; Qin, W.; Moeck, P.; Mandell, E.: Making sense of nanocrystal lattice fringes. J. Appl. Phys. 98 (2005) 114308-1-114308-10; arXiv:cond-mat/0212281 v2; Virtual Journal of Nanoscale Science and Technology Vol. 12 (2005) Issue 25.

[8] Moeck, P.; Zahornadský, J.; Dušek, B.; Fraundorf, P.: Image-based Nanocrystallography with on-line Database Support, Proc. of SPIE Vol. 6370 (2006) 63701A-1-63701A-12, October 1-4, 2006, eds. N. K. Dhar, A.K. Dutta, and M. Islam.

[9] Bjorge, R.: Lattice-Fringe Fingerprinting: Structural Identification of Nanocrystals Employing High-Resolution Transmission Electron Microscopy, MSc thesis, Portland State University, May 9, 2007.

[10] Moeck, P.; Bjoerge, R.; Mandell, E.; Fraundorf, P.: Lattice-fringe fingerprinting of an iron-oxide nanocrystal supported by an open-access database. Proc. NSTI-Nanotech Vol. 4 (2007) 93-96, (www.nsti.org, ISBN 1-4200637-6-6).

[11] Moeck, P.; Seipel, B.; Bjorge, R.; Fraundorf, P.: Lattice fringe fingerprinting in two dimensions with database support. Proc. NSTI-Nanotech Vol. 1 (2006) 741-744, (www.nsti.org, ISBN 0-9767985-6-5).

[12] Moeck, P.; Certík, O.; Seipel, B.; Groebner, R.; Noice, L.; Upreti, G.; Fraundorf, P.; Erni, R.; Browning, N. D.; Kiesow, A.; Jolivet, J.-P.: Identifying unknown nanocrystals by fringe fingerprinting in two dimensions & free-access crystallographic databases. Proc. of SPIE Vol. 6000 (2005) 60000M-1-60000M-12.

[13] Leslie M. (Ed.): Free the Crystals. Science 310 (2005) 597.

[14] Le Bail, A.: Does open data better serve the crystallographic community? International Union of Crystallography Newsletter 12(2) (2004) 27, www.iucr.org, ISSN 1067-0696.

[15] Chateigner, D.; Chen, X.; Ciriotti, M.; Cranswick, L. M. D.; Downs, R. T.; Le Bail, A.; Lutterotti, L.; Yokochi, A. F. T.: COD (Crystallography Open Database) and PCOD (Predicted). In: Abstracts XX[th] Congress of the International Union of Crystallography (IUCr), Florence (Italy), August 23 - 31, 2005.

[16] http://crystallography.net

[17] http://nanocrystallography.research.pdx.edu/CIF-searchable/COD.php

[18] Hall, S.; McMahon, B. (editors): International Tables for Crystallography, Vol. G: Definition and exchange of crystallographic data. International Union of Crystallography, Chester 2005. http://www.iucr.org/iucr-top/cif/index.html

[19] Brown, I. D.; McMahon, B.: CIF: the computer language of crystallography. Acta Cryst. B 58 (2002) 317-324.

[20] Spence, J. C. H.; Wu, J. S.: Electron nanocrystallography - solving small crystal structures by TED. Microsc. Microanal. 11(Suppl 2) (2005) 532-532.

[21] Zuo, J. M.: Electron nanocrystallography. In: Handbook of Microscopy for Nanotechnology (Eds. N. Yao, Z. L. Wang), p. 567-600, Kluwer Academic Publ., Boston, Dordrecht, New York, London, 2005.

[22] Hovmöller, S.; Sjögren, A.; Farrats, G.; Sundberg, M.; Marinder, B.-O.: Accurate atomic positions from electron microscopy. Nature 311 (1984) 238-241.

[23] Sjögren, A.: Crystallographic Electron Microscopy. Chem. Commun. 8 (1987), Doctoral Dissertation, Stockholm University, ISSN 91-7146-533-7.

[24] Zuo, X. D.; Hovmöller, S.; Electron Crystallography: Structure Determination by HREM and Electron Diffraction. In: Industrial Applications of Electron Microscopy (Ed. Z. R. Li), p. 583-614. Marcel Dekker Inc., 2003.

[25] Zuo, X. D.: Electron Crystallography of Inorganic Structures – Theory and Practice. Chem. Commun. 5 (1995), Doctoral Dissertation, Stockholm University, ISSN 0366-5607.

[26] Carlsson, A.; Oku, T.; Bovin, J.-O.; Wallenberg, R.; Malm, J.-O.; Schmid, G.; Kubicki, T.: The First Structure Determination of Nanosized Colloidal Particles of $Pd_3P$ by High-Resolution Electron Microscopy. Angew. Chem. Int. Ed. 37 (1998) 1217-1220

[27] Ohsuna, T.; Liu, Z.; Terasaki, O.; Hiraga, K.; Camblor M. A.: Framework Determination of a Polytype of Zeolite Beta by Using Electron Crystallography. J. Phys. Chem. B 106 (2002) 5673-5678.

[28] Dorset, D. L.: Correlations, convolutions and the validity of electron crystallography. Z. Kristallogr. 218 (2003) 237-246.

[29] Dorset, D. L.: Structural Electron Crystallography. Plenum Press, New York and London 1995.

[30] Dorset, D. L.; Hovmöller, S.; Zuo, X. D. (Eds.): Electron Crystallography. Kluwer Academic Publ. 1997.

[31] Weirich, T. E.; Lábár, J. L.; Zuo, X. D. (Eds.): Electron Crystallography, Novel Approaches for Structure Determination of Nanosized Materials. Springer, 2006.

[32] Gjønnes, J.; Hansen, V.; Berg, B. S.; Runde, P.; Cheng, Y. F.; Gilmore, C. J.; Dorset, D. L.: Structure model for the phase $Al_mFe$ derived from three-dimensional electron diffraction intensity data collected by a precession technique. Comparison with convergent beam diffraction. Acta Cryst. A 54 (1998) 306-319.

[33] Gemmi, M.; Zuo, X. D.; Hovmöller, S.; Migliori, A.; Vennström, M.; Andersson, Y.: Structure of $Ti_2P$ solved by three-dimensional electron diffraction data collected with the precession technique and high-resolution electron microscopy. Acta Cryst. A 59 (2003) 117-126.

[34] Gjønnes, J.; Hansen, V.; Kverneland, A.: The Precession Technique in Electron Diffraction and its Application to Structure Determination of Nano-Size Precipitates in Alloys. Microsc. Microanal. 10 (2004) 16-20.

[35] Own, C. S.; Sinkler, W.; Marks, L. D.: Rapid structure determination of a metal oxide from pseudo-kinematical electron diffraction data. Ultramicroscopy 106 (2006) 114-122.

[36] Dorset, D. L.: The crystal structure of ZSM-10, a powder X-ray and electron diffraction study. Z. Kristallogr. 221 (2006) 260-265.

[37] Fraundorf, P.: Stereo analysis of single crystal electron diffraction data. Ultramicroscopy 6 (1981) 227-235.

[38] Fraundorf, P.: Stereo analysis of electron diffraction from known crystals. Ultramicroscopy 7 (1981) 203-205.

[39] Fraundorf, P.: Determining the 3D lattice parameters of nanometer-sized single crystals from images. Ultramicroscopy 22 (1987) 225-229.

[40] Mighell A. D.; Himes, V. L.: A new method for phase identification for electron diffractionists. J. Electr. Microsc. Techn. 16 (1990) 155-159.

[41] Qin, W.; Fraundorf, P.: Lattice parameters from direct-space images at two tilts. Ultramicroscopy 94 (2003) 245-262; arXiv:cond-mat/0001139

[42] Lábár, J. L: Consistent indexing of a (set of) SAED pattern(s) with the ProcessDiffraction program. Ultramicroscopy 103 (2005) 237-249.

[43] Moeck, P.; Fraundorf, P.: Image-based Nanocrystallography in three dimensions by means of Transmission Electron Goniometry with free on-line Database Support. In: Materials and Systems, Proc. Materials Science and Technology (MS&T), Vol. 2 (2006) 533-543; arXiv:cond-mat/0611345

[44] Hovmöller, S.: PhIDO – Phase identification and indexing in seconds, Calidris, Manhemsvägen 4, SE-191 45 Sollentuna, Sweden, see footnote no.[3].

[45] Le Page, Y.: Accurate d-spacings from zero-order Laue zone patterns. Microsc. Res. Techn. 23 (1992) 243-247.

[46] Carr, M. J.; Chambers, W. F.; Melgaard, D.: A Search/Match Procedure for Electron Diffraction Data Based on Pattern Matching in Binary Bit Maps. Powder Diffraction 1 (1986) 226-234.

[47] Hart, H. V.: ZONES: a search/match database for single-crystal electron diffraction. J. Appl. Cryst. 35 (2002) 552-555.

[48] Lyman, C. E.; Carr, M. J.: Identification of Unknowns in Electron Diffraction. In: Electron Diffraction Techniques (Ed. J. M. Cowley), Vol. 2, p. 373 - 417, Oxford University Press 1992.

[49] Vincent, R.; Midgley, P.: Double conical beam-rocking system for measurement of integrated electron diffraction intensities. Ultramicroscopy 53 (1994) 271-282

[50] Avilov, A.; Kuligin, K.; Nicolopoulos, S.; Nickolskiy, M.; Boulahya, K.; Portillo, J.; Lepeshov, G.; Sobolev, B.; Collette, J. P.; Martin, N.; Robins, A. C.; Fischione, P.: Precession technique and electron diffractometry as new tools for crystal structure analysis and chemical bonding determination. Ultramicroscopy 107 (2007) 431-444.

[51] Oleynikov, P.; Hovmöller, S.; Zuo, X. D.: Precession electron diffraction: Observed and calculated intensities. Ultramicroscopy 107 (2007) 523-533.

[52] Gjønnes, K.; Cheng, Y. F.; Berg, B. S.; Hansen, V.: Corrections for multiple scattering in integrated intensities. Application to determination of structure factors in the (001) projection of $Al_mFe$. Acta Cryst. A 54 (1998) 102-119.

[53] Morniroli, J. P.; Redjaïmia, A.; Nicolopoulos, S.: Contribution of electron precession to the identification of the space group from microdiffraction patterns. Ultramicroscopy 107 (2007) 514-522.

[54] Mighell, A. D.; Karen, V. L.; NIST Crystallographic Databases for Research and Analysis. J. Res. Natl. Inst. Stand. Technol. 101 (1996) 273-280; NIST Standard Reference Database 3, http://www.nist.gov/srd/nist3.htm

[55] Carr, M. J.; Chambers, W. F.; Melgaard, D.; Himes, V. L.; Stalick, J.; Mighell, A. D.: NIST/Sandia/ICDD Electron Diffraction Database: A database for phase identification by electron diffraction. J. Res. Natl. Inst. Stds. Technol. 94 (1989) 15-20; NIST Standard Reference Database 15, http://www.nist.gov/srd/nist15.htm

[56] Ayer, R.: Determination of unit cell. J. Electron. Microsc. Technique 13 (1989) 16-26.

[57] Le Page, Y.; Downham, D. A.: Primitive unit cell volumes obtained from unindexed convergent-beam electron diffraction patterns. J. Electron. Microsc. Technique 18 (1991) 437-439.

[58] Le Page, Y.: Ab-initio primitive cell parameters from single convergent-beam electron diffraction patterns: A converse route to the identification of microcrystals with electrons. Microsc. Res. Technique 21 (1992) 158-165.

[59] Zuo, J. M.: New method of Bravais lattice determination. Ultramicroscopy 52 (1993) 459-464.

[60] Kim, G.-H.; Kim, H.-S.; Kum, D.-W.: Simple procedure for phase identification using convergent beam electron diffraction patterns. Microsc. Res. Techn. 33 (1996) 510-515.

[61] Zuo, J. M.; Kim, M.; Holmestad, R.: A new approach to lattice parameter measurements using dynamic electron diffraction electron diffraction and pattern matching. J. Electron Microsc. 47 (1998) 121-127.

[62] Denley, D. R.; Hart, H. V.: RINGS: a new search/match database for identification by polycrystalline electron diffraction. J. Appl. Cryst. 35 (2002) 546-551.

[63] Làbàr, J. L.: Phase identification by combining local composition from EDX with information from diffraction database. In: Electron Crystallography: Novel Approaches for Structure Determination of Nanosized Materials (Eds. T. E. Weirich, J. L. Làbàr, X. D. Zuo), p. 207-218, Springer 2006. (Proc. of the NATO Advances Study Institute on Electron Crystallography, Erice, Italy, 10-14 June 2004).

[64] Weirich, Th. E.; Winterer, M.; Seifried, S.; Hahn, H.; Fuess, H.: Rietveld analysis of electron powder diffraction data from nanocrystalline anatase, $TiO_2$. Ultramicroscopy 81 (2000) 263-270.

[65] Young, R. A., (Ed.): The Rietveld Method. Oxford University Press 1995.

[66] Pawley, G. S.: Unit cell refinement form powder diffraction scans. J. Appl. Cryst. 14 (1981) 357-361.

[67] Le Bail, A.; Duroy, H.; Fourquet, J. L.: Ab initio structure determination of $LiSbWO_6$ by X-ray powder diffraction. Mat. Res. Bull. 23 (1988) 447-452.

[68] Sivia, D. S; David, W. I. F.: A Bayesian-approach to extracting structure-factor amplitudes from powder diffraction data. Acta Cryst. A 50 (1994) 703-714.

[69] Le Bail, A.: Whole powder pattern decomposition methods and applications: A retrospection. Powder Diffraction 20 (2005) 316-326.

[70] Vainshtein, B. K.; Zvyagin, B. B.; Avilov, A. S.: Electron Diffraction Structure Analysis. In: Electron Diffraction Techniques (Ed. J. M. Cowley), Vol. 1, p. 216-312, Oxford University Press 1992.

[71] Eades, J. A.: Convergent-Beam Diffraction. In: Electron Diffraction Techniques (Ed. J.M. Cowley), Vol. 1, p. 313-359, Oxford University Press, New York 1992.

[72] Mansfield, J.: Practical phase identification by convergent beam electron diffraction. J. Electr. Microsc. Techn. 13 (1989) 3-15.

[73] Amelinckx, S.: III.4. Electron Diffraction in Transmission Electron Microscopy. In: Fifty Years of Electron Diffraction (Ed. P. Goodman), p. 378-396, D. Reidel Publishing Company 1981.

[74] Krivanek, O. L.: Practical High-resolution electron microscopy. In: High-Resolution Transmission Electron Microscopy and Associated Techniques (Eds. P. R. Buseck, J. M. Cowley, and L. Eyring), p. 519-567, Oxford University Press 1988.

[75] Allpress, J. G.; Hewat, E. A.; Moodie, A. F.; Sanders, J. V.: n-beam lattice images. I. Experimental and computed images from $W_4Nb_{26}O_{77}$. Acta Cryst. A 28 (1972) 528-536.

[76] Kirkland, E. J.: Advanced computing in electron microscopy. Plenum, New York 1998.

[77] Stadelmann, P.: EMS - A software package for electron-diffraction analysis and HREM image simulation in materials science. Ultramicroscopy 21 (1988) 131-145; Electron Microscopy Image Simulation - EMS On Line, CIME-EPFL; http://cimesg1.epfl.ch/CIOL/; jems, the EMS java version; http://cimewww.epfl.ch/people/stadelmann/jemsWebSite/jems.html

[78] Zuo, J. M.; Mabon, J. C.: Web Electron Microscopy Applications Software (WebEMAPS), University of Illinois at Urbana – Champaign, http://emaps.mrl.uiuc.edu/

[79] Du, K.; von Hochmeister, K.; Phillipp, F.: Quantitative comparison of image contrast and pattern between experimental and simulated high-resolution transmission electron micrographs. Ultramicroscopy 107 (2007) 281-292.

[80] ctfExplorer by M. V. Sidorov, http://clik.to/ctfexplorer

[81] The Bristol Group under the direction of John Steeds: Convergent Beam Electron Diffraction of Alloy Phases, Adam Hilger Ltd. 1984.

[82] http://icsdweb.fiz-karlsruhe.de/index.php *and* http://icsd.ill.fr/icsd/index.html and http://www.stn-international.de/stndatabases/databases/icsd.html

[83] http://www.crystalimpact.com/ *and* http://www.crystalimpact.com/pcd/download.htm

[84] http://crystdb.nims.go.jp/

[85] Klug, A.: Image Analysis and Reconstruction in the Electron Microscopy of Biological Macromolecules. Chem. Scr. 14 (1978-1979) 245-256.

[86] Hovmöller, S.: CRISP: crystallographic image processing on a personal computer. Ultramicroscopy 41 (1992) 121-135.

[87] Li, F. H.; Tang, D.: Pseudo-weak-phase-object approximation in high-resolution electron microscopy I. Theory. Acta Cryst. A 41 (1985) 376-382.

[88] Tang, D.; Teng, C. M.; Zuo, J.; Li, F. H.: Pseudo-weak-phase-object approximation in high-resolution electron microscopy II. Feasibility of directly observing $Li^+$. Acta Cryst. B 42 (1986) 340-342.

[89] Self, P. G.; Bhadeshia, H. K. D. H.; Stobbs, W. M.: Lattice spacings from lattice fringes. Ultramicroscopy 6 (1981) 29-40.

[90] de Ruijter, W. J.; Sharma, R.; McCartney, M. R.; Smith, D. J.: Measurement of lattice-fringe vectors from digital HREM images: experimental precision. Ultramicroscopy 57 (1995) 409-422.

[91] Biskupek J.; Kaiser, U.: Practical considerations on the determination of the accuracy of the lattice parameter measurements from digital recorded diffractograms. J. Electron. Microsc. 53 (2004) 601-610.

[92] de Ruijter, W. J.: Determination of lattice parameters from digital micrographs based on measurements in reciprocal space. J. Comp. Assist. Microsc. 6 (1994) 195-212.

[93] Jillavenkatesa, A.; Kelly, J. F.: Nanopowder characterization: challenges and future directions. J. Nanopart. Res. 4 (2002) 463-468.

[94] Pennycook, S. J.; Varela, M.; Hetherington, C. J. D.; Kirkland, A. I.: Materials advances through aberration-corrected electron microscopy. MRS Bulletin 31 (January 2006) 36-43.

[95] Wang, P., Bleloch, A. L.; Falke, U.; Goodhew, P. J.: Geometric aspects of lattice contrast visibility in nanocrystalline materials using HAADF STEM. Ultramicroscopy 106 (2006) 277-283.

[96] Own, C. S.; Sinkler, W.; Marks, L. D.: Prospects for aberration corrected electron precession. Ultramicroscopy 107 (2007) 534-542.

[97] Lu, K.; Zhao, Y. H.: Experimental evidences of lattice distortion in nanocrystalline materials. NanoStruct. Mater. 12 (1999) 559-562.

[98] Sui, M. L.; Lu, K: Variations in lattice parameter with grain size of a nanophase Ni3P compound. Mater. Sci. Engin. A 179/180 (1994) 541-544.

[99] Roduner, E.: Nanoscopic Materials, Size-dependent Phenomena. Royal Society of Chemistry, Cambridge 2006.

[100] Mighell, A. D.: The Normalized Reduced Form and Cell Mathematical Tools for Lattice Analysis - Symmetry and Similarity. J. Res. Natl. Inst. Stand. Technol. 108 (2003) 447-452.

[101] Andrews, L. C.; Bernstein, H. J.; Pelletier, G. A.: A Perturbation Stable Cell Comparison Technique. Acta Cryst. A 36 (1980) 248-252.

[102] Pielaszek, R.; Lojkowski, W.; Matysiak, H.; Wejrzanowski, T.; Opalinska, A.; Fedyk, R.; Burjan, A.; Proykova, A.; Iliev, H.: In: Eighth Nanoforum Report: Nanometrology, (Eds. W. Lojkowski, R. Turan, A. Proykova, A. Daniszewska), p. 79-80, July 2006, freely available at www.nanoforum.org

[103] Banfield, J. F.; Zhang, H.: Nanoparticles in the Environment. In: Nanoparticles and the Environment (Eds. J. F. Banfield, A. Navrotsky), p. 1-58, Reviews in Mineralogy & Geochemistry, Vol. 44, Mineralogical Society of America, (series editor: P. H. Ribbe) 2001.

[104] Mitome, M.: Visibility of Si nanoparticles embedded in an amorphous $SiO_2$ matrix. J. Electr. Microsc. 55 (2006) 201-207.

[105] Petkov, V.; Ragan, K. K.: Kanatzidis, M. G.; Billinge, S. J. L.: Structure f crystallographically challenged materials by profile analysis of atomic pair distribution functions: Study of $LiMoS_2$ and mesostructured $MnGe_4S_{10}$. Mat. Res. Soc. Symp. Proc. Vol. 678 (2001) EE1.5.1-EE1.5.6.

[106] Billinge, S. J. L.; Kanatzidis, M. G.: Beyond crystallography: the study of disorder, nanocrystallinity, and crystallographically challenged materials with pair distribution functions. Chem. Commun. 749 (2004) 749-760.

[107] Billinge, S. J. L.; Levin, I.: The problem with Determining Atomic Structure at the Nanoscale, Science 316 (2007) 561-565.

[108] Sakamoto, Y.; Kaneda, M.; Terasaki, O.; Zhao, D. Y.; Kim, J. M.; Shin, H. J.; Ryoo, R.: Direct imaging of the pores and cages of three-dimensional mesoporous materials. Nature 408 (2000) 449-453.

[109] http://nanocrystallography.research.pdx.edu/CIF-searchable/NCD.php

[110] Brodsky, M. H.: Fair and Useful Copyright. ACA RefleXions, Newsletter of the American Crystallographic Association, Number 4, 2006, p. 15-16 (ISSN 1958-9945), http://aca.hwi.buffalo.edu//newsletterpg_list/Newsletters_PDF/Winter06.pdf

[111] http://www.wipo.int/treaties/en/ip/berne/trtdocs_wo001.html

[112] http://www.wipo.int/treaties/en/ip/berne/pdf/trtdocs_wo001.pdf

[113] http://www.iucr.org/iucr-top/cif/faq/

**Footnotes**

[1] The crystal thickness limits up to which the scattering of fast electrons by a nanocrystal may be treated by the kinematic or quasi-kinematic[2] diffraction approximations varies from sample to sample and with the instrumental parameters. Factors that influence these limits are, for example, the atomic number of the major scattering components, the crystal structure, and the acceleration voltage of the microscope. No definitive thickness limits can, therefore, be given that would apply to all nanocrystals. A crystal thickness below about 5 to 10 nm may, however, suffice in many cases for the kinematic or quasi-kinematic[2] diffraction approximations to be applicable. Monographs (e.g. Vainshtein, B. K.: Structure Analysis by Electron Diffraction, Pergamon Press Ltd., Oxford, 1964) and reviews (e.g. Humphreys, C. J., The scattering of fast electrons by crystals. Rep. Prog. Phys. 42 (1979) 1827-1887) give frequently values in this range as approximate limits for electrons that have been accelerated by a potential of more than 100 kV. Higher potentials and crystals structures that consist of light atoms are likely to increase these limits.
Over more than the last six decades, there have been theoretical analyses of the limiting cases for which the dynamical electron diffraction theory may for sufficiently thin crystals be approximated with its kinematical or quasi-kinematical[2] counterparts, e.g. Blackman, M.: On the intensities of electron diffraction rings, Proc. Royal Society London A 173 (1939) 68-82. Due to the more recent experimental success of electron crystallography, the analysis of these limiting cases with novel theoretical approaches has become an active research field in its own right, see footnote no.[2]. "Blackman's correction for two-beam dynamical diffraction effects" can, for example, now be used for the analysis of partially crystalline (nano-)materials (Peng, L.-M., Dudarev, S. L., Whelan, M. J.: High-Energy Electron Diffraction and Microscopy, Oxford University Press 2003) and in conjunction with electron precession diffraction data[3] for crystals that are several tens of nanometer thick (Own, C. S.; Subramanian, A. K.; Marks, L. D.: Quantitative Analyses of Precession Diffraction Data for a Large Cell Oxide. Microsc. Microanal. 10 (2004) 96-104 and Christopher Su-Yan Own's PhD thesis, System Design and Verification of the Precession Electron Diffraction Technique, Northwestern University, 2005, freely downloadable at: http://www.numis.northwestern.edu/Research/Current/precession.shtml and Sinkler, W.; Own, C. S.; Marks, L. D.: Application of a 2-beam model for improving the structure factors from precession electron diffraction intensities. Ultramicroscopy 107 (2007) 543-550).
Based on the combination of electron diffraction and HRTEM imaging data, an empirical method has been demonstrated for the partial correction of distortions of electron diffraction intensities that are due to the Ewald-sphere curvature, crystal bending, crystal thickness variations and other causes (Huang, D. X.; Liu, W.; Gu, Y. X.; Xiong, J. W.; Fan H. F.; Li F. H.: A Method of Electron Diffraction Intensity Correction in Combination with High-Resolution Electron Microscopy. Acta Cryst. A 52 (1996) 152-157). That method also reduces multiple dynamical scattering effects (Lu, B.; Li, F. H.; Wan, Z. H.; Fan, H. F.; Mao, Z. Q.: Electron crystallographic study of $Bi_4(Sr_{0.75}La_{0.25})_8Cu_5O_y$ structure. Ultramicroscopy 70 (1997) 13-22).

[2] It has been shown that for the scattering of fast electrons by thin crystals with little or negligible overlap between the electrostatic potential of adjacent atoms, one may use complex "dynamical scattering factors" instead of their well known real kinematic atomic scattering factor counterparts in kinematic-type expressions for the calculation of dynamical diffracted-beam amplitudes. When overlaps in the electrostatic potential can no longer be neglected, i.e. typically in cases of thicker crystals, the dynamical scattering factors may in similar kinematic-type expressions be replaced by complex "dynamical string scattering factors" (where the string parameters are the atomic species and the number of atoms in the string). Extensive computations revealed that the proposed kinematic-type approximations may even work for crystals with thicknesses much larger than 10 nm (Peng, L.-M.; Quasi-dynamical electron diffraction – a kinematic type of expression for the dynamical diffracted-beam amplitudes. Acta Cryst. A 56 (2000) 511-518, see also Peng, L.-M., Dudarev, S. L., Whelan, M. J.: High-Energy Electron Diffraction and Microscopy, Oxford University Press 2003).
This goes to show that the experimental success of (quasi-)kinematic-theory-based electron crystallography leads in turn to theoretical developments in its support. Because (quasi-) kinematic-theory-based electron crystallography works for inorganic crystals in the 5 to 10 nm thickness range, the much less sophisticated structural fingerprinting on the basis of (quasi-)kinematic-theories should work as well. Note that there is also an electron channeling approximation to dynamic diffraction for moderately low acceleration voltages and moderately heavy atomic columns (such as 200 kV and Cu in the [001] orientation, Van Dyck, D.; Op de Beeck, M.: A simple intuitive theory for electron diffraction. Ultramicroscopy 64 (1996) 99-107 and Van Aert, S.; Geuens, P.; Van Dyck, D.; Kisielowski, C.; Jinschek, J. R.: Electron channelling based crystallography. Ultramicroscopy 107 (2007) 551-558) that supports the extraction of the projected electrostatic potential from HRTEM images for crystal thicknesses up to about 10 nm.

While it is known that a single Cu atom does scatter fast electrons dynamically, a calculated [001] „kinematic structure map" of $C_{32}N_8Cl_{16}Cu$ (copper hexadecachlorophthalocyanine) for 10.7 nm thickness, 400 kV, and all scattering vectors out to 0.1 nm$^{-1}$ represented the projected electrostatic potential quite faithfully as a result of channeling effects and because many of the low-indexed diffracted beams behaved kinematically. The *"range of crystal thicknesses for which single-crystal structure analysis is feasible is"*, thus, *"not limited to the range of validity of the kinematic approximation."*, Peng, L.-M.; Wang, S. Q.: On the Validity of the Direct Phasing and Fourier Method in Electron Crystallography, Acta Cryst. A 50 (1994) 759-771. Structural fingerprinting of nanocrystals from HRTEM images may, therefore, not be limited to the range of validity of the kinematic approximation either. (Electron crystallography from inorganic crystals in the thickness range 10-20 nm has also been demonstrated on the basis of the electron channeling theory, Sinkler, W.; Marks L. D.: Dynamical direct methods for everyone. Ultramicroscopy 75 (1999) 251-268).

It has also been demonstrated experimentally that the "effective thicknesses" of 15 to 20 nm sized inorganic nanocrystals can (in a variety of orientations for multiple crystals) be reduced to levels for which electron crystallography was feasible (despite multiple scattering effects) by excluding the central parts of HRTEM images from the analysis, e.g. Carlsson, A.; Oku, T.; Bovin, J.-O.; Wallenberg, R.; Malm, J.-O.; Schmid, G.; Kubicki, T.: The First Structure Determination of Nanosized Colloidal Particles of $Pd_5P$ by High-Resolution Electron Microscopy. Angew. Chem. Int. Ed. 37 (1998) 1217-1220. Corresponding image simulations by these authors for this material showed that multiple diffraction effects start to affect the amplitudes of the Fourier coefficients of the projected electrostatic potential for thicknesses that exceed 5 to 7 nm. Assuming spherical nanocrystals, the "maximal effective thicknesses" can, however, in this study be approximated to exceed 10 nm.

[3] PhIDO, ELD and CRISP run on IBM compatible personal computers under Windows$^{TM}$. These programs are part of a comprehensive software suite for electron crystallography, have been developed by Sven Hovmöller and coworkers, and can be purchased form the following Swedish company: Calidris, Manhemsvägen 4, SE-191 45 Sollentuna, Sweden, Telephone +46 8 16 23 80 (daytime in Europe) +46 8 96 77 22 (weekends and European evenings), Fax +46 8 625 00 41, http://www.calidris-em.com/ and http://www.fos.su.se/~svenh/. The crystallography database file for the PhIDO program comes with 77 entries from the NIST/NBS database and can be amended and edited with a simple text editor. The ELD program actually performs the least squared fits to the position of either all or selected electron diffraction spots (Zuo, X. D., Sukharev, Y.; Hovmöller, S.: ELD – a computer program system for extracting intensities form electron diffraction patterns. Ultramicroscopy 49 (1993) 147-158 and Zuo, X. D., Sukharev, Y.; Hovmöller, S.: Quantitative electron diffraction – new features in the program system ELD. Ultramicroscopy 52 (1993) 436-444).

There is also the electron crystallography freeware program EDM: Electron Direct Methods, by Roar Kilaas, Chris Own, Bin Deng, Kenji Tsuda, Wharton Sinkler, and Laurence Marks. That program is freely downloadable for a range of computer platforms (including IBM compatible personal computers) at http://www.numis.northwestern.edu/edm/ and an earlier version of it is described in Kilaas, R.; Marks, L. D.; Own, C. S.: EDM 1.0: Electron direct methods. Ultramicroscopy 102 (2005) 233-237.

[4] The electron precession method is formally analogous to the well known X-ray precession technique (Buerger, M. J.: Contemporary crystallography. McGraw-Hill, 1970, p. 149-185), but utilizes a precession movement of the electron beam around the microscopes optical axis rather than that of the specimen goniometer around a fixed beam direction. The diffracted beams are de-scanned in such a manner that stationary diffraction patterns are obtained. The illuminating beam can be either parallel or focused. A pseudo-two beam correction to the (quasi-kinematic) intensities might be employed by assuming that these intensities are proportional to the amplitudes of the corresponding structure factors.

Electron precession add-ons to TEMs have been developed by Stavros Nicolopoulos and coworkers and can be purchased from the following Belgian company: NanoMEGAS SPRL, Boulevard Edmond Machterns No 79, Sint Jean Molenbeek, Brussels, B-1080, Belgium, Telephone: +34 649 810 619, http://www.nanomegas.com, info@nanomegas.com.

[5] The beam convergence in convergent beam electron diffraction (CBED) effectively expands the diffraction spots into diffraction disks with crystallographically distinct fine structure. Structural information in 3D can be extracted for a given goniometer setting by using higher order Laue zone information in CBED patterns. Because multiple

beam diffraction conditions are well defined in CBED, structure factor amplitudes can be extracted on the basis of the dynamical diffraction theory.

[6] The term structure image has been proposed by John M. Cowley to describe a member of the restricted set of lattice images in TEM which can be directly interpreted to some limited resolution in terms of a crystal's projected atomistic structure. Structure images need to be obtained under instrumental conditions which are independent of the crystal structure. Weak phase objects that are imaged at the Scherzer (de)focus fulfill this condition. (Spence, C. H.: High-Resolution Electron Microscopy, 3[rd] edition, Appendix 4, p. 391, Oxford University Press 2004).

[7] STEM images typically show scan-aberration artifacts such as distortions in the resolved spatial frequencies, resulting in distortions in the angles between lattice fringes. These distortions can, in principle, be corrected for each individual STEM operated under its typical imaging conditions. (Sanchez, A. M.; Galindo, P. L.; Kret, S.; Falke, M.; Beanland, R.; Goodhew, P. J.: An approach to the systematic distortion correction in aberration-corrected HAADF images, J. Microsc. 221 (2006) 1-7).